# Practical quantum cryptography for secure free-space communications

Richard J. Hughes, William T. Buttler, Paul G. Kwiat, Steve K. Lamoreaux,
George L. Morgan, Jane E. Nordholt and C. Glen Peterson

University of California, Los Alamos National Laboratory, Los Alamos, NM 87545, USA

**ABSTRACT**

Quantum cryptography is an emerging technology in which two parties may simultaneously generate shared, secret cryptographic key material using the transmission of quantum states of light. The security of these transmissions is based on the inviolability of the laws of quantum mechanics and information-theoretically secure post-processing methods. An adversary can neither successfully tap the quantum transmissions, nor evade detection, owing to Heisenberg's uncertainty principle. In this paper we describe the theory of quantum cryptography, and the most recent results from our experimental free-space system with which we have demonstrated for the first time the feasibility of quantum key generation over a point-to-point outdoor atmospheric path in daylight. We achieved a transmission distance of 0.5 km, which was limited only by the length of the test range. Our results provide strong evidence that cryptographic key material could be generated on demand between a ground station and a satellite (or between two satellites), allowing a satellite to be securely re-keyed on orbit. We present a feasibility analysis of surface-to-satellite quantum key generation.

## 1. INTRODUCTION

Two of the main goals of cryptography (encryption and authentication of messages) can be accomplished, with provable security, if the sender ("Alice") and recipient ("Bob") possess a secret random bit sequence known as "key" material. The initial step of key distribution, in which the two parties acquire the key material, must be accomplished with a high level of confidence that a third party ("Eve") cannot acquire even partial information about the random bit sequence. If Alice and Bob communicate solely through classical messages it is impossible for them to generate a certifiably secret key owing to the possibility of passive eavesdropping. However, secure key generation becomes possible if they communicate with single-photon transmissions using the emerging technology of quantum cryptography, or more accurately, quantum key distribution (QKD).[1] (A small amount of shared secret key material is required to perform initial authentication.)

The security of QKD is based on the inviolability of the laws of quantum mechanics and provably secure (information theoretic) public discussion protocols. Eve can neither "tap" the key transmissions owing to the indivisibility of quanta[2] nor copy them faithfully because of the quantum "no-cloning" theorem.[3] At a deeper level, QKD resists interception and retransmission by an eavesdropper because in quantum mechanics, in contrast to the classical world, the result of a measurement cannot be thought of as revealing a "possessed value" of a quantum state. A unique aspect of quantum cryptography is that Heisenberg's uncertainty principle ensures that if Eve attempts to intercept and measure Alice's quantum transmissions, her activities must produce an irreversible change in the quantum states (she "collapses the wavefunction") that are retransmitted to Bob. These changes will introduce an anomalously high error rate in the transmissions between Alice and Bob, allowing them to detect the attempted eavesdropping. In particular, from the observed error rate Alice and Bob can put an upper bound on any partial knowledge that an eavesdropper may have acquired by monitoring their transmissions. This bound allows the intended users to apply conventional information theoretic techniques by public discussion to distill an error-free, secret key.

Because it has the ultimate security assurance of a law of Nature, quantum cryptography offers potentially attractive "ease of use" advantages over conventional key distribution schemes: it avoids the "insider threat" because key material does not exist before the quantum transmissions take place; it replaces cumbersome conventional key distribution methods whose security is based on the physical security of the distribution process; and it provides a secure alternative to key distribution schemes based on public key cryptography, which are potentially vulnerable to algorithmic advances and improved computing techniques. Thus, quantum key distribution enables "encrypted communications on demand," because it allows key generation at transmission time over an unsecured optical communications link.

The first quantum key distribution protocol was published by Charles Bennett and Gilles Brassard in 1984 and is now known as "BB84".[4] A further advance in theoretical quantum cryptography took place in 1991 when Ekert proposed[5] that Einstein-Podolsky-Rosen (EPR) "entangled" two-particle states could be used to implement a quantum cryptography protocol whose security was based on Bell's inequalities. Starting in 1989, Bennett, Brassard and collaborators performed the first experimental demonstration of QKD by constructing a working prototype system for the BB84 protocol, using polarized photons.[6] Although the propagation distance was only about 30 cm, this experiment is in several ways still the most thorough demonstration of quantum cryptography.

Potentially practical applications of QKD, outside the carefully controlled environment of a physics laboratory, are largely determined by the physics of single-photon production, the requirement of faithful transmission of the quantum states involved, the existence of high-efficiency single-photon detectors at the required wavelengths, and the compatibility of QKD with existing optical communications infrastructures. In 1992 Bennett published a "minimal" QKD scheme ("B92") and proposed that it could be implemented using single-photon interference with photons propagating for long distances over optical fibers.[7] Since then, several experimental groups [8, 9, 10, 11] have developed optical fiber-based QKD systems. For example, at Los Alamos we have demonstrated the feasibility of low-error rate QKD over underground optical fibers that were installed for network applications.[11] We have previously demonstrated QKD over 24 km of fiber[12] and have operated for over one year at an increased propagation distance of 48 km.[13] In recent years there have also been considerable developments in the use of free-space laser communications[14] for high-bandwidth terrestrial, surface-to-satellite, satellite-to-satellite and (potentially) deep-space communications. The optical pointing, acquisition and tracking techniques developed for laser communications could be used to make QKD possible over line-of-sight transmissions in free-space,[15, 16, 17] provided that signal-to-noise and bit rates adequate for cryptographic applications can be achieved. There are certain key distribution problems for which free-space QKD would have definite practical advantages. For example, it is impractical to send a courier to a satellite. We believe that free-space QKD could be used for key generation between a low-earth orbit satellite and a ground station,[17] as well as in other applications where laser communications are possible. To demonstrate this possibility we have developed a free-space QKD system for such applications and have previously achieved a transmission distance of 1-km over a folded path (to a reflector and back) at night.[17] More recently, we have performed the first demonstration of free-space QKD over a point-to-point 0.5-km path in daylight, and we report these results here.

The remainder of this paper is organized as follows. In section 2 we give a concise introduction to the theory of quantum cryptography. Then, in section 3 we describe the experimental considerations underlying our implementation of quantum cryptography in our free-space QKD system. In section 4 we present a feasibility study of QKD between a ground station and a satellite in low-earth orbit. Finally, in section 5 we present some conclusions.

## 2. QUANTUM CRYPTOGRAPHY: THEORY

To understand QKD we must first move away from the traditional key distribution metaphor of Alice sending *particular* key data to Bob. Instead, we should have in mind a more symmetrical starting point, in which Alice and Bob initially generate their own, independent random binary sequences, containing more numbers than they need for the key material that they will ultimately share. Through public discussion they agree on a QKD protocol by which they can perform a bitwise comparison of their sequences using a quantum transmission (over a "quantum channel") and a public discussion of the results (over an authenticated public channel) to distill a shared, random subsequence, which will become the key material. It is important to appreciate that they do not need to identify *all* of their shared numbers, or even *particular* ones, because the only requirements on the key material are that the numbers should be secret and random. Several QKD protocols have been developed, but for simplicity we shall describe the minimal B92 QKD protocol[7] in terms of the preparation and measurement of single-photon polarization states. (Cryptographically, the BB84 protocol has certain advantages, but the physics issues involved are identical with B92.)

In the B92 QKD protocol Alice can produce photons with either of two non-orthogonal polarizations: V or +45° (say); and Bob can make either of two complementary non-orthogonal polarization measurements: –45° or H (say). Alice and Bob generate their own independent sequences of random binary numbers. Next, they proceed through their sequences bit-by-bit in synchronization, with Alice preparing a polarized photon for each of her bits according to the rules:

$$"0" \leftrightarrow V$$
$$"1" \leftrightarrow +45°$$
(1)

Alice sends each photon over a "quantum channel" to Bob. (The quantum channel is a transmission medium that isolates the quantum state from interactions with the "environment.") Bob makes a polarization measurement on each photon he receives, according to the value of his bit as given by:

$$"0" \leftrightarrow -45°$$
$$"1" \leftrightarrow H$$
(2)

and records the result ("pass" = Y, "fail" = N). Note that Bob will never record a "pass" (a false positive) if his bit is different from Alice's (they have crossed polarizers). He only records a "pass" on 50% of the bits that they have in common. In the example of four bits shown in Figure 1, we see that for the first and fourth bits Alice and Bob had different bit values, so that Bob's result is a definite "fail" in each case. However, for the second and third bits, Alice and Bob have the same bit values and the protocol is such that there is a probability of 0.5 that Bob's result will be a "pass" in each case. Of course, we cannot predict in any particular experiment which one will be a "pass," but in this example the second bit was a "fail" and the third bit was a "pass."

| Alice's bit value | 1 | 0 | 1 | 0 |
| --- | --- | --- | --- | --- |
| Alice's polarization | +45° | V | +45° | V |
| Bob's polarization | –45° | –45° | H | H |
| Bob's bit value | 0 | 0 | 1 | 1 |
| Bob's results | N | N | Y | N |

**Figure 1**. A four-bit example of B92 quantum key distribution

To complete the protocol Bob sends a copy of his (Y or N) *results* to Alice, but not the measurement that he made on each bit. (It is at this data-reconciliation stage that the initial key material is required for authentication. This key material can be replaced by a portion of the key material generated by QKD.) He may send this information over a conventional (public) channel which may be subject to eavesdropping. Now Alice and Bob retain only those bits for which Bob's result was "Y" and these bits become the shared key material. (In the example of Figure 1 the third bit becomes the first bit of the shared key.) An ideal B92 procedure distills on average one shared bit from every four initial bits assuming that there are no photon losses in transmission or detection. The 25% efficiency, $\eta_Q$, of the idealized QKD process is the price that Alice and Bob must pay for secrecy. In a practical system, additional losses in transmission (efficiency factor $\eta_T$) and detection (efficiency factor $\eta_D$) will occur (and can be tolerated). However, these losses only affect the bit rate, not the security.

In a practical system there will be errors in the reconciled data arising from optical imperfections and detector noise, which must be removed before the key material can be used. Alice and Bob can remove these errors using conventional error correcting codes over their public channel, but at the expense of revealing some (parity) information about the resulting key material to Eve. Errors and information leakage will also occur if Eve performs her own measurement of Alice's states on the quantum channel and fabricates new photons to send on to Bob. To take an extreme case, if Eve measures each of Alice's photons using Alice's basis she will introduce a 25% error rate into Alice and Bob's key material, while correctly identifying 75% of Alice's bits. Of course, Alice and Bob could readily detect such a large error rate and would not then use their reconciled data for key material, but the eavesdropper could still gain some information at the expense of a proportionately smaller error rate if she only measures a fraction of Alice's photons. It is the goal of quantum cryptography for Alice and Bob to translate an observed error rate into an upper bound on Eve's knowledge of their reconciled data.[18] Such bounds have been established for eavesdropping attacks on individual bits[19] and are the subject of current research in the case of coherent attacks on multiple bits. Error correction should then be followed by a further stage of "privacy amplification" to reduce any partial knowledge acquired by Eve to less than one bit of the final key string.[20] For example, Alice and Bob could choose the parities of random subsets of their error corrected data so that Eve will be forced to have less than one bit of information about the resulting key. These additional stages are performed over the public channel.

Authentication of the public channel transmissions is necessary to avoid a "man-in-the-middle" attack, in which Eve could gain control of both the quantum and public channels, allowing her to masquerade as Bob to Alice and vice-versa. Alice and Bob would then unknowingly generate independent keys with Eve who could use these keys to read all of their subsequent encrypted communications. Alice and Bob need a short, secret authentication key to start the QKD procedure, and can replenish this key with a small portion of the QKD material generated. For authentication based on random hashing they will need $O(\log_2 n)$ secret authentication bits for every *n*-bit public transmission.[21]

So from the foregoing, we see that a QKD procedure may be broken down into the following seven stages:

1. Alice and Bob acquire a secret authentication key;
2. Alice and Bob generate independent secret sequences of random bits;
3. Alice and Bob use the quantum transmissions of a QKD protocol to compare their sequences and classical communications to identify a random subsequence of shared secret bits;
4. Alice and Bob perform an error correction procedure on the data;
5. Alice and Bob assess (from the error rate) how much knowledge Eve may have acquired;

6. Alice and Bob perform an appropriate privacy amplification procedure over the public channel;
7. Part of the resulting key material is used to replenish the authentication bits required in step 1, so that the system is ready for the next key generation session.

The result of these steps is a shared, error-free secret key. (It has been proposed that the key bits generated by QKD should be used for the encryption of communications using the unbreakable "one-time pad" method.[22] However, the key material could equally well, and more practically, be used by Alice and Bob in any other symmetric key cryptosystem.) Of the steps above, only one (step 3) involves the experimental physics issues that will be crucial to the practical feasibility of QKD. In our work we have therefore focussed our efforts on this component of QKD. A fully functional key generation system would include careful implementation of the other steps, but these (with the exception of step 5) are better understood and may be readily incorporated once step 3 has been adequately demonstrated. Step 5 relates to the physics of eavesdropping and a full treatment of this topic is beyond the scope of this paper. We will therefore limit ourselves to a few additional remarks on this subject.

In the simple form described above, the B92 protocol is vulnerable to Eve measuring Alice's photons in Bob's basis and only sending on those photons she can identify. (A "Bob's basis" attack.) This will cause a factor of four reduction in bit rate unless Eve sends out multiple photons instead of just one. Alice and Bob can protect against this type of attack if Bob is able to detect the photon number of the received bits, as in our system described below. They could also avoid this problem entirely by using the BB84 protocol, which uses four states instead of two. However, from the perspective of the physics, the B92 and BB84 protocols are so similar that BB84 will also be possible under conditions for which QKD with the B92 protocol is feasible.

In considering possible eavesdropping on a QKD system it also important to distinguish between attacks that are possible with existing technology, which are limited to individual bit attacks, and potential future attacks that are limited only by the laws of physics. In particular, current QKD experiments use approximate single-photon states that are obtained by attenuating the output of a pulsed laser so that the average photon number per pulse is less than one. Such pulses contain a Poisson distribution of photon numbers, and the low intensity is necessary to ensure that very few pulses are vulnerable to an eavesdropper using an optical beamsplitter to "tap out" a photon from pulses containing more than one photon. Present QKD system can be made secure against such attacks by appropriate use of privacy amplification. However, as quantum-optical technology advances an eavesdropper could use more sophisticated methods to attack such a system in the future, as we will discuss below. Before such attacks become possible it will be important for Alice and Bob to replace their weak laser pulse QKD source with a true single-photon light source. Several techniques are now becoming feasible for producing such states of light. A demonstration of the feasibility of QKD with weak laser pulses also implies the viability of QKD with a true single-photon light source under the same experimental conditions, because of the linearity of the processes involved.

## 3. EXPERIMENTAL POINT-TO-POINT QUANTUM KEY GENERATION OVER 0.5 KM IN DAYLIGHT

The success of QKD over free-space optical paths depends on the transmission and detection of single optical photons against a high background through a turbulent medium. Although these are challenging problems they can be overcome with careful choices of experimental parameters and the use of various optical techniques developed for laser communications. The atmosphere has a high transmission "window" for light with a wavelength in the vicinity of 770 nm. Photons can be readily produced at this wavelength with rugged, low-power semiconductor lasers and their polarization properties controlled with off-the-shelf optical components. Furthermore, commercial single-photon counting modules (SPCMs) are now available that can count such photons with efficiencies as high as $\eta_D \sim 65\%$ at rates of up to 1 MHz, with dark count rates as low as 50 Hz. The atmosphere is essentially non-birefringent at these wavelengths and so will allow the faithful transmission of the QKD polarization states. However, atmospheric turbulence will introduce both photon arrival time jitter and beam wander (through variations in refractive index). The slow turbulence time-scales involved (0.1s to 0.01s) allow the jitter to be compensated by transmitting a bright timing laser pulse (which carries no key information) at a different wavelength a short time (100 ns, say) before each QKD photon. The arrival of this bright pulse at the receiver allows a definite timing window to be imposed for the single QKD photon's arrival, because the atmospheric transmission time will not have changed over the intervening short interval. Beam wander caused by atmospheric turbulence reduces the QKD bit rate, but as we will see later is not a critical limitation on surface-to-satellite paths even if left uncontrolled. However, active beam steering ("tip-tilt" control) methods have been developed for laser communications to keep the beam directed onto the receiver. For example, by monitoring a reflected component of the bright timing pulse, an error signal can be derived and fedback to a beam-steering mechanism.

At first sight a more serious concern is that the large background of photons from the sun (or even the moon at night) could swamp the single-photon QKD signal. However, as we will see below, a combination of (sub)-nanosecond timing, narrow wavelength filters[23, 24] and a small solid angle for photon acceptance (spatial filtering) at the receiver[16] can render this background tractable.

The QKD transmitter ("Alice") in our system contains a 1-MHz clock that synchronizes the various events. (See Figure 2.)

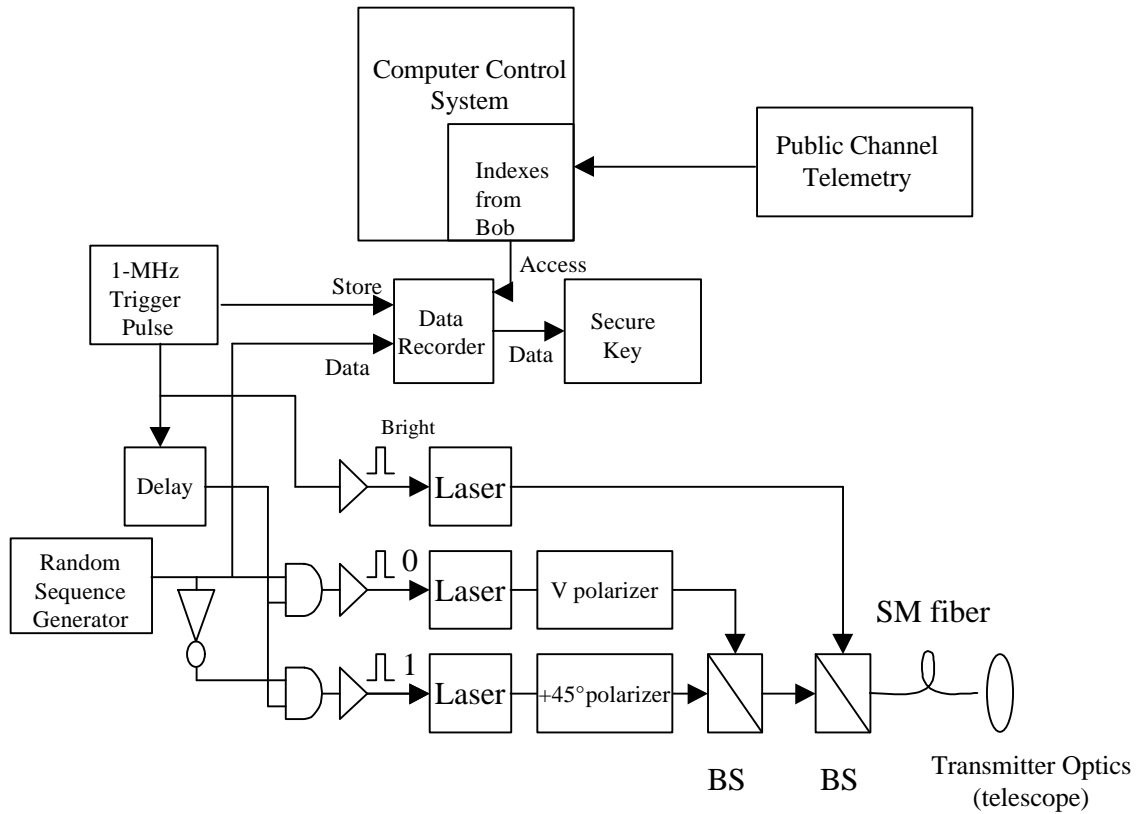

**Figure 2**. QKD transmitter ("Alice") block diagram.

On each "tick" of the clock a ~ 1-ns optical "bright pulse" is produced from a "timing-pulse" laser operating at a wavelength of ~ 768 nm. After a ~ 100-ns delay one of two temperature-controlled "data" diode lasers emits a ~ 1-ns optical pulse that is attenuated to the single-photon level and constrained by interference filters to a wavelength of 773.0±0.5 nm. The polarization of the optical pulse from each laser is set to one of the two non-orthogonal settings required for the B92 protocol. The choice of which data laser fires is determined by a random bit value that is obtained by discriminating electrical noise. The random bit value is indexed by the clock tick and recorded in a computer control system's memory. All three optical pulse paths are combined (using beam splitters, BS), directed into a single-mode (SM) optical fiber for delivery to a transmitting telescope, and emitted towards Bob's receiver. The process is then repeated one microsecond later with the next random bit, and so on.

At Bob's QKD receiver the light pulses are collected by a 3.5-inch diameter Cassegrain telescope and directed into a polarization analysis and detection system. (See Figure 3.)

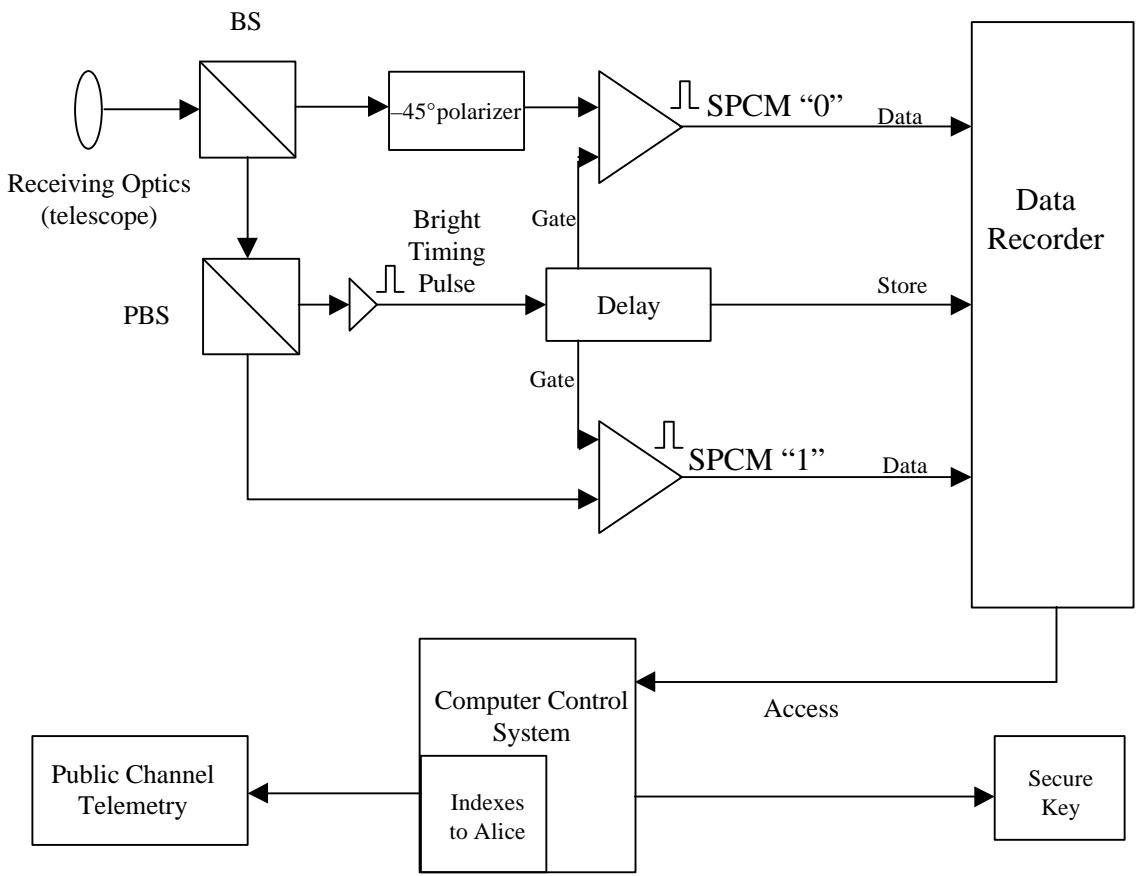

**Figure 3**. QKD receiver ("Bob") block diagram.

The bright pulse triggers an avalanche photodiode detector, and this event sets up an electronic timing "window" about 5-ns long in which a QKD optical data pulse is expected. After emerging from the collection telescope, an optical data pulse encounters an optical beamsplitter at which a single-photon would be either transmitted or reflected with equal probabilities. We use this quantum-mechanically random behavior at the beamsplitter to perform Bob's random choice of which B92 polarization measurement is made on the arriving optical data pulse. Along the transmitted path, an optical data pulse's polarization is analyzed according to Bob's B92 "0" value, while along the reflected path a measurement for *H*-polarization is made using a polarizing beamsplitter (PBS). (The PBS transmits *H*- but reflects *V*-polarization.) After each polarization analysis stage, optical data pulses pass through interference filters matched to those in the transmitter, and are collected into (spatial filtering) multi-mode optical fibers for delivery to single-photon counting modules (SPCMs), one for each bit value. Of course, for many of the arriving bright pulses there will be no corresponding single-photon detection owing to the efficiency of the B92 protocol, the attenuation experienced by the optical data pulses, and the SPCM's detection efficiency. For events on which one of the two SPCMs triggers, Bob can assign a bit value to Alice's transmitted bit. He records these detected bits in the memory of a computer control system, indexed by the "bright pulse" clock tick. Subsequently, Bob's computer control system transmits a file of index values (but *not* the corresponding bit values) to Alice over a wireless Ethernet link. Alice and Bob then use those detected bits as the raw bit sequences from which an error-free, secret key is distilled using further communications over the Ethernet channel.

The QKD system was operated for several days over a 0.5-km horizontal outdoors atmospheric range from west (transmitter) to east (receiver) under daylight and nighttime conditions. A typical sample of 256 bits identified from 50,000 initial bits under daylight conditions on November 19, 1998 at 4.30pm, is shown in Figure 4, with Alice's bit value marked as "A" and Bob's as "B".

A 10101110 10011001 10001111 **0**0001101 10011011 10110011 10011001 10**0**00001
B 10101110 10011001 10001111 **1**0001101 10011011 10110011 10011001 10**1**00001

A 11001110 01101010 10010111 10001110 10110000 11001110 11101101 10110011
B 11001110 01101010 10010111 10001110 10110000 11001110 11101101 10110011

A 0100000**0** 01010001 00010000 00010010 01000111 00010011 11001000 010**0**1001
B 0100000**1** 01010001 00010000 00010010 01000111 00010011 11001000 010**1**1001

A **0**1110101 10110010 01110010 11101111 00101101 00010101 10011111 00101111
B **1**1110101 10110010 01110010 11101111 00101101 00010101 10011111 00101111

**Figure 4**. A raw sample of bits identified by Alice and Bob as shared ones, using the B92 protocol over a 0.5-km daylight path.

The above data set contains 5 errors (marked in bold type) and the bit error rate (BER) observed in the whole data set was approximately 1.6%. This would be regarded as unacceptably high in any conventional telecommunications application, but can be tolerated in QKD because of the secrecy of the bits. The effectiveness of our precise timing, wavelength and spatial filtering techniques for mitigating daylight background photon events is shown by the measured background rate of 1 event per 50,000 detector triggers, contributing only approximately 0.4% to the BER. Detector dark noise makes an even smaller contribution of approximately 0.1% to the BER. We conclude that the dominant contribution to the BER is from optical misalignment and intrinsic imperfections of the polarizing elements.

Clearly, errors must be removed before the bit strings can be used as key material. An efficient, interactive error correction procedure has been invented that can remove *all* errors from such data sets, with BERs of up to 15%.[25] However, for simplicity in our system we perform a two-dimensional block-parity error correction procedure over the Ethernet channel, which requires Alice to reveal some parity data about the bit strings. An eavesdropper could combine this information with any knowledge acquired through eavesdropping on the quantum transmissions. There are two ways of dealing with this issue. Alice and Bob could encrypt the parity information, which would require them to have more secret bits initially, or they could perform additional privacy amplification to compensate for the information revealed, which would produce a shorter key string. We perform a rudimentary privacy amplification procedure by dropping one row and one column from each matrix of data bits. A fully functional QKD system would incorporate a more sophisticated privacy amplification procedure.

The ~ 5-kHz key rate is adequate for the one-time pad encryption of small image files that we have incorporated into our software control system. Because the one-time pad method requires as many key bits as message bits, the key rate would not be adequate for more lengthy transmissions. This key rate would be acceptable and better used for generating session keys for use in other symmetric key cryptosystems because such keys need only be a few hundred bits in length.

The average photon number per optical data pulse for this data set was ~ 0.3, giving a probability of 22% that the pulse contains exactly one photon, and a probability of 25.9% that a pulse contains at least one photon. Thus, approximately 15% of the detectable pulses contain more than one photon. (Such multi-photon pulses can trigger *both* of Bob's SPCMs, but the rate for these "dual fire" errors is reduced below the key rate by the product of the BER and the multi-photon emission probability. We observed no dual fires in the entire 50-k bit sequence leading to the data in Figure 4. By monitoring the dual-fire rate, Alice and Bob could protect against the "Bob's basis" attack outlined in Section 2.) So, a full security analysis of our system must take into account the possibility of Eve performing a beamsplitting attack to "tap off" the occasional photon from two-photon pulses. (See ref.[17] for an analysis of this type of attack.) With appropriate privacy amplification procedures our system can be rendered secure against this and other individual bit attacks that are possible with existing technology. However, in the future a system such as ours could become vulnerable to a so-called QND attack,[26, 18] in which Eve uses a quantum non-demolition (QND) measurement to identify those pulses containing two photons. She could then determine Alice's bit value on these pulses, suppress the other pulses, and transmit a new photon to Bob, using a hypothetical lossless channel. Because Alice's two-photon emission rate is larger than Bob's detection rate in our system, Bob would not notice a reduction in bit rate in this type of attack. Although the QND attack is not feasible today, this possibility should not be ignored. We plan eventually to remove this potential vulnerability by using a true single-photon light source instead of a weak pulsed laser source.

**4. QUANTUM KEY DISTRIBUTION TO SATELLITES**

Our proof-of-concept QKD demonstrations over horizontal terrestrial paths provide strong evidence that surface-to-satellite QKD will be possible. This is because the optical influence of turbulence is the major hurdle to be overcome in surface-to-satellite QKD, and the turbulent effects occur predominantly within the lowest 2 km of the atmosphere. Ground-to-satellite, satellite-to-ground and satellite-to-satellite QKD should all be possible, for both low-earth orbit (LEO) and geostationary satellites. For illustration we will here estimate the key generation capability of QKD between a ground station and a LEO satellite (~ 300 km altitude) in one overhead pass (duration ~ 8 minutes). Our objective will be to produce multiple new cryptovariables, each of several hundred bits in length. We will assume that the QKD transmitter (Alice) is at the ground station and the receiver (Bob) is on the satellite. (Similar arguments support the viability of satellite-to-ground QKD transmissions, which would have key rate and hardware advantages.)

We have designed our QKD system to operate at a wavelength near 770 nm where the atmospheric transmission from surface to space can be as high as 80%. Furthermore, at optical wavelengths the polarized QKD photons can be faithfully transmitted because the depolarizing effects of Faraday rotation in the ionosphere are negligible. Because the atmosphere is only weakly dispersive, a bright timing pulse (which carries no key information) of ~ 100-ps duration can be used to set a short time window (~ 1 ns) within which to look for the QKD photon. A single QKD-photon arriving ~ 100 ns after the bright pulse would find that the satellite had moved by less than 1 mm.

To estimate the rate at which QKD photons would be detected at the satellite from the ground station transmitter, we assume 20-cm diameter optics at both the transmitter and satellite receiver, leading to a ~ 1-m diameter diffraction-limited spot size at a 300-km altitude satellite. However, there will be beam-wander owing to atmospheric turbulence, which at night in a high desert location such as Los Alamos can be 1 to 5 arc seconds.[27] For this analysis we assume a worst case "seeing" of ~ 10 times the diffraction limit (i.e. 10 arc seconds of wander) so that the photon collection efficiency at the satellite is ~ $10^{-4}$. Thus, with a laser pulse rate of 10 MHz, one photon-per-pulse on average, an atmospheric transmission of ~ 80%, a 65% detector efficiency and allowing for the 25% intrinsic efficiency of the B92 QKD protocol, a key generation rate of ~ 250 Hz is feasible. (There would be a factor of two higher key rate with the BB84 protocol.)

Higher key rates would be possible under more typical seeing conditions. Also, with a simple beam tilt feedback system, as used in laser communications systems, the beam could be locked onto the satellite, increasing the key rate to ~ 40 kHz. A retro-reflector on the satellite would return a portion of each bright pulse to the transmitter with a ~ 2 ms delay, which is much shorter than the time-scale of atmospheric turbulence fluctuations. (From the ground, the satellite would move through an angle of only ~ 50 micro radians in this time.) It would also be possible to place the QKD transmitter on the satellite and the receiver on the ground. Because most of the optical influence of atmospheric turbulence would occur in the final ~ 2-km of the beam path, a higher key rate would then be possible even without tilt control.

To determine if this key rate is useful we must also consider the error rate. We first consider errors arising from background photons arriving at the satellite on a nighttime orbit with a full moon and under (poor) 10-arc second seeing conditions. A typical radiance observed at the satellite at the transmission wavelength would then be ~ 1 mW $m^{-2}$ $str^{-1}$ $\mu m^{-1}$ or ~ $4 \times 10^{15}$ photons $s^{-1}$ $m^{-2}$ $str^{-1}$ $\mu m^{-1}$. We will assume that the receiver "sees" a solid angle ~ five times the apparent size of the source (i.e. 5 arc seconds) and that there is a 1-nm bandwidth interference filter placed in front of the detector, giving a background photon arrival rate of ~ 225 Hz (full moon). (For comparison, detector dark counts would be ~ 50 Hz.) However, the single-photon detector would only be triggered by precursor bright pulses impinging on the satellite, giving a detector trigger rate of ~ 90 kHz (without beam tilt control). With a 1-ns time window applied to the detector, the (fractional) bit error rate (BER) from background photons would therefore be ~ $5 \times 10^{-5}$ (full moon). With beam tilt control the fractional BER from background photons would be ~ $4 \times 10^{-5}$. In practice, errors from optical component limitations and misalignments will be larger, amounting to a 1 to 2 percent BER based on our experience.

From this simple analysis using worst-case estimates, we see that QKD between a ground station and a low-earth orbit satellite should be possible on nighttime orbits. During the several minutes that a satellite would be in view of the ground station there would be adequate time to acquire the satellite, perform the QKD transmissions for ~ 1 minute, and produce a minimum of ~ 10,000 raw bits, from which a shorter error-free key stream of several thousand bits would be produced after error correction and privacy amplification. Under more typical seeing conditions or with beam tilt control implemented, up to $10^5$ key bits could be produced in the 1-minute QKD transmission. A cryptographically useful quantity of key material could therefore be generated between a ground station and a LEO or geostationary satellite using available technology. (Satellite to satellite QKD transmissions would also be possible.)

On daytime orbits the background radiance would be ~ 4,000 times larger (~ $2 \times 10^{19}$ photons $s^{-1}$ $m^{-2}$ $str^{-1}$ $\mu m^{-1}$) than under a full moon, but a narrow atomic vapor filter (~ $10^{-2}$ nm filter width) [28] would keep the background photon arrival rate to only ~ 10 kHz. Assuming a typical daytime seeing of 10 arc seconds,[27] the key rate would be ~ 250 Hz, and the BER from background photons would then be ~ $2 \times 10^{-3}$ (without tilt control). QKD is therefore also likely to be possible on daytime orbits.

# 5. SUMMARY AND CONCLUSIONS

This paper presents the first practical demonstration that point-to-point free-space QKD is feasible under daylight conditions outside a laboratory, achieving a realistic propagation distance of 0.5 km that was only limited by the length of the test range. We are now in the process of improving the system and anticipate performing a 2-km daylight demonstration early in 1999, possibly increasing to 7 km later. Free-space QKD could therefore be used in conjunction with terrestrial laser communications systems that are now commercially available. Our results also provide strong evidence that cryptographic key material could be generated on demand between a ground station and a satellite (or between two satellites) using QKD, allowing a satellite to be securely re-keyed on orbit.

The development of QKD for satellite communications would represent a major step forward in both security and convenience. If the key material supplied at launch should be used up during normal operations or compromised, an issue arises of how to securely re-key a satellite on-orbit. In contrast to conventional key distribution methods whose security is based on assumptions of computational complexity, QKD is a physics-based technique and as such needs to be experimentally validated under the conditions of its intended use. To our knowledge the primary physics requirements for this application of QKD, namely the transmission and detection of single photons between a ground station and an orbital asset, have never been demonstrated. However, many of the optical acquisition, pointing, tracking and adaptive optics techniques developed for laser communications with satellites can be directly applied to this problem. Therefore, we believe that a surface-to-satellite QKD demonstration experiment would be a logical and realistic next step in the development of this new field. Furthermore, we believe that the development of QKD for re-keying of satellites on-orbit would be prudent, so as to have an alternative to traditional key distribution methods that can potentially become vulnerable to unanticipated algorithmic or computational advances.

Satellite QKD could also be used to provide secure key distribution to two ground-based users (Alice and Bob) who do not have access to optical fiber communications and who are not within line-of-sight: they could each generate independent quantum keys with the same satellite, which would then transmit the XOR of the keys to Bob. Bob would then XOR this bit string with his key to produce a key that agrees with Alice's. Alice and Bob could then use their shared key for encrypted communications over any convenient channel. This procedure could extend the security and convenience of QKD to widely separated ground-based users.

# ACKNOWLEDGEMENTS

It is a pleasure to thank Charles Bennett, Norbert Lutkenhaus, J. David Murley, John Rarity and Robert Shea for helpful conversations. We especially thank Kevin McCabe for preparing Figure 2.